# Two conjectures such that the proof of any one of them will lead to the proof that P = NP

Malay Dutta (Tezpur University India)

## ABSTRACT

In this paper we define a construct called a time-graph. A complete time-graph of order n is the cartesian product of a complete graph with n vertices and a linear graph with n vertices.  A time-graph of order n is given by a subset of the set of edges E(n) of such a graph. The notion of a hamiltonian time-graph is defined in a natural way and we define the Hamiltonian time-graph problem (HAMTG) as : Given a time-graph is it hamiltonian ? We show that the Hamiltonian path problem (HAMP) can be transformed to HAMTG in polynomial time. We then define certain vector spaces of functions from E(n) and E(n) x E(n) to B = {0,1}, the field of two elements and derive certain properties of these spaces. We give two conjectures about these spaces and prove that if any one of these conjectures is true, we get a polynomial time algorithm for the Hamiltonian path problem. Since the Hamiltonian path problem is NP-complete we obtain the proof of P = NP provided any one of the two conjectures is true.

## 1. Introduction

The P = NP problem is one of the main open problems in Theoretical Computer Science today. The classes P, NP, NP-complete and NP-hard problems are defined in (1). P = NP can be proved by constructing a polynomial time algorithm for any of the NP-complete or NP-hard problems. The Hamiltonian path problem is : Given a graph, does it have a hamiltonian path, that is a path which traverses every vertex exactly once ? By corollary 1 of Theorem 15.6 of (1), this problem is NP-complete. In this paper we give two conjectures such that if any one of them is proved we can prove the existence of a polynomial time algorithm for this problem and hence P = NP.

Since Khachian (2), gave a polynomial time algorithm for linear programming, most of the earlier efforts to prove P = NP were to obtain a polynomial time reducttion of some NP-hard optimization problem for example the traveling salesman problem (TSP) to linear programming. In section 2, we elaborate on such an effort and discuss how the ideas presented in this paper evolved from this.

In section 3, we formally define the construct time-graph which is based on the notion of the time-dependent traveling salesman problem (3). We also define hamiltonian time graphs and the Hamiltonian time graph problem. We define the vector spaces $H^n$ and $H_P^n$ and derive certain properties of these spaces.

In section 4, we propose the two conjectures and prove the main consequences which are proved assuming any one of the conjectures to be true.

In section 5, we prove certain results which leads to a polynomial time algorithm for the construction of a basis of $H_P^n$.

In section 6, we present a polynomial time algorithm for the Hamiltonian path problem and prove its correctness provided any one of the conjectures proposed in this paper is true and thereby conclude that under this condition P = NP.

## 2. An earlier effort to prove P = NP

Most of the earlier efforts to prove P = NP were to obtain polynomial time reduction of NP-hard optimization problems to linear programming since linear programming has a polynomial time algorithm as proved by Khachian (2). An important effort of this type is described in section 2.1 of (4). Let E be a finite ground set and I be a set of subsets of E. With every element e of E we associate a variable $x_e$ that is a component of a vector x in $R^E$ indexed by e. With every subset F of E, we associate a vector $x^F$ in $R^E$, defined as

$$x^F_e = 1 \quad \text{if e is in F , 0 otherwise}$$

We take $P_I$ to be the convex hull of $x^F$'s for F in I. Now suppose every e in E is associated with a weight $c_e$. For every F in E we can now define a cost function $c(F) = \sum_{e \in F} c_e$. Then we can solve the combinatorial optimization problem of optimizing c(F) over F in I, by solving the linear programming problem of optimizing $c^T x$ over the polytope $P_I$. In order to apply linear programming techniques we need a complete description of the polytope $P_I$ by way of linear equations and inequalities. However such a completeness result has proved completely elusive for NP-hard problems like the traveling salesman problem.

In this connection it may be worthwhile to investigate a polytope obtained from vectors $y^F$ in $R^{E \times E}$ defined as

$$Y^F_{(e,e')} = 1 \quad \text{if both e, e' are in F, 0 otherwise}$$

We then take $AP_I$ to be the intersection of the affine space generated by the $y_F$'s with the non-negative orthant of $R^{E \times E}$. Taking the linear map P from $R^{E \times E}$ to $R^E$ given by

$$(Py^F)_e = y^F_{(e,e)}$$

we can hope to obtain $P_I$ as the image of $AP_I$ under this map. Since the polytope $AP_I$ is polynomially described, we can get a possible polynomial time reduction of our optimization problem to linear programming.

Following this approach, in this paper we replace the field R of real numbers by the field B = {0,1} of two elements. A reduction of the Hamiltonian path problem to linear programming gets replaced by the reduction to the solution of linear equations in a polynomial number of variables, provided any one of the conjectures proposed in the paper is true.

## 3. Preliminaries for the work presented in this paper

A complete time-graph of order n, $K_T^n$ is a layered graph with vertices (i,t) for i,t=1,2,..n. The index t represents the layer number. Each edge (i,j,t) for i,j = 1,2,…,n and t=1,2,…,n-1 connects (i,t) and (j,t+1). Thus $K_T^n$ is nothing but the direct product of $K^n$ the complete graph of n vertices and the linear graph of n vertices. The set of edges of $K_T^n$ will be denoted by E(n). A time-graph of order n will be a subgraph of $K_T^n$ with the same set of vertices as that of $K_T^n$ and the set of edges E a subset of E(n). A time-graph will therefore be denoted by the set of its edes E.

Let $S_n$ denote the set of permutations of {1,2,…,n}. An edge (i,j,t) in $K_T^n$ is said to be incident on a permutation $\pi \in S_n$ if $\pi(t) = i$ and $\pi(t+1) = j$. A permutation $\pi \in S_n$ is said to be incident on a time-graph G of order n if every edge incident on $\pi$ is in G. G is said to be Hamiltonian if there is at least one permutation incident on it. The Hamiltonian time-graph problem (HAMTG) is : Given a time-graph, is it Hamiltonian ?

The Hamiltonian path problem (HAMP) is : Given a graph, does it have a hamiltonian path, that is a path which traverses every vertex exactly once ? This problem is known to be NP-complete (Corollary 1 of Theorem 15.6 of (1)). The following is easy to prove.

**Theorem 1 :** HAMP can be transformed to HAMTG in $O(n^3)$ time.

Proof : Given a graph G = (V,E) with the set of vertices V = {1,2,…,n} and the set of edges E, we construct a time-graph $G_T$ of order n as follows :

For every t = 1,2,…,n-1 (i,j,t) is in $G_T$ if and only if {i,j} is in E.

Suppose G has a hamiltonian path ($\pi(1), \pi(2),…, \pi(n)$). Then $\pi \in S_n$ and ($\pi(t),\pi(t+1)$) is in E for t=1,2,…n-1. Therefore ($\pi(t),\pi(t+1),t$) is in $G_T$ and hence $\pi$ is incident on $G_T$. Therefore $G_T$ is hamiltonian.

Conversely if $G_T$ is hamiltonian, let $\pi \in S_n$ be incident on $G_T$. Then $(\pi(t),\pi(t+1),t)$ is in $G_T$ that is $(\pi(t),\pi(t+1))$ is in E for t=1,2,…,n-1. Therefore $\pi$ gives a hamiltonian path in G.

Hence G has a hamiltonian path if and only if $G_T$ is hamiltonian. Also the construction of $G_T$ can obviously be carried out in $O(n^3)$ time. This proves the theorem.

Let B = {0,1} be the field of two elements. We define the mappings $T^n : S_n \to B^{E(n)}$ and $T_p^n : S_n \to B^{E(n) \times E(n)}$ by

$T^n(\pi)(e) = 1$ if e is incident on $\pi$, 0 otherwise

and $\quad T_p^n(\pi)(e,e') = 1$ if both e and e' are incident on $\pi$, 0 otherwise

Note that both $B^{E(n)}$ and $B^{E(n) \times E(n)}$ are vector spaces over B under pointwise operations. We define $H^n$ to be the linear span of $T^n(S_n)$ and $H_p^n$ to be the linear span of $T_p^n(S_n)$. We now define a mapping P from $H_p^n$ to $B^{E(n)}$ by

$P(g)(e) = g(e,e)$

Obviously P is linear and for any g in $T_p^n(S_n)$ and hence for any g in $H_p^n$, $g(e,e') = g(e',e)$. We also have

**Theorem 2 :** For any $\pi$ in $S_n$, $P(T_p^n(\pi)) = T^n(\pi)$.

Proof : $\quad\quad\quad\quad\quad\quad T^n(\pi)(e) = 1$

If and only if $\quad\quad$ e is incident on $\pi$

If and only if $\quad\quad T_p^n(\pi)(e,e) = 1$

If and only if $\quad\quad P(T_p^n(\pi))(e) = 1$

Hence $\quad\quad T^n(\pi)(e) = P(T_p^n(\pi))(e)$ for all e and the result follows.

**Corollary 2.1 :** P is an onto map from $H_p^n$ to $H^n$.

Proof : This follows from Theorem 2 and the facts that (i) $H_p^n$ is the linear span $T_p^n(S_n)$ (ii) $H^n$ is the linear span of $T^n(S_n)$ and (ii) P is linear.

We also define a mapping $P_e$ for any e in E(n) from $H_p^n$ to $B^{E(n)}$ as follows :

$P_e(g)(e') = g(e,e')$. $\quad$ Obviously $P_e$ is linear and $P_e(g)(e') = P_{e'}(g)(e)$. We also have

**Theorem 3 :** For any $\pi$ in $S_n$ $P_e(T_p^n(\pi)) = T^n(\pi)$ if e is incident on $\pi$, 0 otherwise.

Proof : If e is incident on $\pi$, for any e' in E(n)

$P_e(T_p^n(\pi))(e') = 1$

If and only if  $T_p^n(\pi)(e,e') = 1$

If and only if $e'$ is incident on $\pi$

If and only if  $T^n(\pi)(e') = 1$  and hence $P_e(T_p^n(\pi)) = T^n(\pi)$

If $e$ is not incident on $\pi$  $P_e(T_p^n(\pi))(e') = T_p^n(\pi)(e,e') = 0$ and hence $P_e(T_p^n(\pi)) = 0$

This proves the theorem.

**Corollary 3.1 :** $P_e$ is a mapping from $H_p^n$ to $H^n$.

Proof : This follows from Theorem 3 and the facts that (i) $P_e$ is linear (ii) $H_p^n$ is the linear span of $T_p^n(S_n)$ and (iii) $H^n$ is the linear span of $T^n(S_n)$.

For any $f$ in $H^n$ we define the value of $f$, $v(f)$ by

$$v(f) = \sum_{i,j} f(i,j,1)$$

For $g$ in $H_p^n$ we define $v(g) = v(P(f))$

Clearly $v$ is linear and we also have

**Theorem 4 :** For any $f$ in $T^n(S_n)$, $v(f) = 1$

Proof : Let $f = T^n(\pi)$. Then in the definition of $v(f)$ exactly one term for which $i = \pi(1)$ and $j = \pi(2)$ will be 1 and hence $v(f) = 1$.

**Corollary 4.1 :** For any $g$ in $T_p^n(S_n)$, $v(g) = 1$

Proof : Let $g = T_p^n(\pi)$. Then $v(g) = v(P(T_p^n(\pi)))$

$\qquad\qquad\qquad\qquad = v(T^n(\pi))$ \qquad using Theorem 2

$\qquad\qquad\qquad\qquad = 1$ \qquad\qquad using Theorem 4

The following are now obvious.

**Corollary 4.2 :** If $f = \sum_{i=1,k} f_i$ with $f_i$ in $T^n(S_n)$ then $v(f) = 1$ if and only if $k$ is odd.

**Corollary 4.3 :** If $g = \sum_{i=1,k} g_i$ with $g_i$ in $T_p^n(S_n)$ then $v(g) = 1$ if and only if $k$ is odd.

**Corollary 4.4 :** If $f = \sum_{i=1,k} \alpha_i f_i$ with $f_i$ in $T^n(S_n)$ then $v(f) = \sum_{i=1,k} \alpha_i$ .

**Corollary 4.5 :** If $g = \sum_{i=1,k} \alpha_i g_i$ with $g_i$ in $T_p^n(S_n)$ then $v(g) = \sum_{i=1,k} \alpha_i$ .

**Theorem 5 :** If $f = \sum_{i=1,k} T^n(\pi_i)$ and $e$ is incident on each $\pi_l$ , then $v(f) = f(e)$ .

Proof : Since $e$ is incident on each $\pi_l$ , $T^n(\pi_i)(e) = 1$. Therefore

$$f(e) = \sum_{i=1,k} T^n(\pi_i)(e) = \sum_{i=1,k} 1 = 1 \text{ if and only if } k \text{ is odd}$$

But $v(f) = 1$ if and only if $k$ is odd using Corollary 4.2. This proves the theorem.

Similarly using Corollary 4.3 we can prove

**Theorem 6 :** If $g = \sum_{i=1,k} T_p^n(\pi_i)$ and $e$ is incident on $g$ for every $i$, then $v(g)=g(e,e)$.

An element $f$ in $H^n$ will be called a cycle if $v(f) = 0$ and an element $g$ in $H_p^n$ is called a cycle if $v(g) = v(P(g)) = 0$. An element $g$ in $H_p^n$ is called a closed cycle if $P(g) = 0$. Obviously sum of cycles are cycles and sum of closed cycles are closed cycles.

Let $G$ be a time-graph of order $n$. $S_n(G)$, $H^n(G)$ and $H_p^n(G)$ will denote the set of permutations incident on $G$, the linear span of $T^n(S_n(G))$ and the linear span of $T_p^n(S_n(G))$ respectively. Let $e_1, e_2, \ldots, e_k$ be an enumeration of $G^c$ the complement of $G$. Let $G_i$ for $i=0, 1, 2, \ldots, k$ be the time-graph $G \cup \{e_j\}_{j=1,2,\ldots,l}$. Thus $G_0 = G$ and $G_k = E(n)$.

We can construct a basis $B = \{f(i,j)\}_{i=0,1,2,\ldots,k;j=1,2,\ldots,d(i)}$ for $H^n$ where $B^{(l)} = \{f(i,j)\}_{i=0,\ldots l; j=1,\ldots d(i)}$ is a basis of $H^n(G_l)$. We shall take each $f(i,j) = T^n(\pi_{ij})$ for some $\pi_{ij}$ in $S_n(G_i)$ and for $i=1,2,\ldots,k$; $e_i$ is incident on $\pi_{ij}$ for any $j=1,2,\ldots,d(i)$. $F(i,j)$ will denote $T_p^n(\pi_{ij})$ and therefore by Theorem 2, $P(F(i,j)) = f(i,j)$. Such a basis can be constructed as follows. We take any basis $B^{(0)} = \{f(0,j)\}_{j=1,2\ldots d(0)}$ of $H^n(G_0)$ with $f(0,j) = T^n(\pi_{0j})$ for $\pi_{0j}$ in $S_n(G_0)$. After obtaining $B^{(l)}$ we extend it to the basis $B^{(l+1)}$ of $H^n(G_{l+1})$ by adding the elements $\{f(l+1,j)\}_{j=1,2,\ldots d(l+1)}$ where $f(l+1,j) = T^n(\pi_{l=1\,j})$. Obviously then $\pi_{l+1\,j}$ is in $S_n(G_{l+1})$ but not in $S_n(G_l)$ i.e. $e_{l+1}$ is incident on $\pi_{l+1\,j}$ for any $j$. This is done for $l = 0, 1, 2, \ldots, k-1$ to get the required basis.

Obviously $d(0) = \dim(H^n(G))$ and $d(i) = \dim(H^n(G_i)) - \dim(H^n(G_{i-1}))$ (may be zero). Such a basis will be called a canonical basis of $H^n$ with respect to $G$ and the enumeration $(e_1, e_2, \ldots, e_k)$ of $G^c$.

Similarly we can construct a basis $B_P = \{g(i,j)\}_{i=0,1,\ldots k;j=1,2,\ldots,c(i)}$ of $H_p^n$ canonical with respect go $G$ and the enumeration $(e_1, e_2, \ldots, e_k)$ of $G^c$. Here we shall take $g(i,j) = T_p^n(\pi`_{ij})$ for some $\pi`_{ij}$ in $S_n$ such that for $i=1,2,\ldots,k$; $\pi`_{ij}$ is incident on $G_i$, and $e_i$ is incident on $\pi`_{ij}$ for any $j = 1,2,\ldots,c(i)$.

Let $g$ be in $H_p^n$ and $P(g) = \sum_{i,j} \alpha(i,j) f(i,j)$. Consider $g_c = g + \sum_{i,j} \alpha(i,j) F(i,j)$. Then $P(g_c) = P(g) + \sum_{i,j} \alpha(i,j) f(i,j) = P(g) + P(g) = 0$. Thus $g_c$ is a closed cycle. Thus we get

**Theorem 7 :** Let $g$ be in $H_p^n$ and $P(g) = \sum_{i,j} \alpha(i,j) f(i,j)$. Then $g = g_c + \sum_{i,j} \alpha(i,j) F(i,j)$ where $g_c$ is a closed cycle.

Let $g$ be in $H_p^n$. An edge $e$ in $E(n)$ is said to support $g$ if $g(e,e') = 1$ for some $e'$ in $E(n)$. The set of edges that support $g$ will be called the support of $g$ (support($g$)). $G$ will be said to be supported in G if support(g) is contained in G, that is $g(e_m, e') = 0$ for all $e'$ and $m = 1, 2, \ldots, k$. where $(e_1, e_2, \ldots, e_k)$ is an enumeration of $G_C$.

For the rest of the section G will be a time-graph of order n, $(e_1, e_2, \ldots, e_k)$ an enumeration of $G^c$, $\{f(i,j)\}_{i=0,1,\ldots,k; j=1,\ldots,d(i)}$ with $f(i,j) = T^n(\pi_{ij})$ a basis of $H^n$ canonical with respect to G and this enumeration of $G^C$ and g an element of $H_P^n$ supported in G. Let $g = \sum_{i,j} \alpha(i,j) F(i,j) + g_c$ where $g_c$ is a closed cycle. Define $f^{(i)} = \sum_{j=1,2,\ldots d(i)} \alpha(i,j) f(i,j)$. We then have

**Theorem 8 :** $\sum_{i \geq m} f^{(i)}(e_m) = 0$ for $m = 1, 2, \ldots, k$.

Proof : We have $P(g)(e_m) = g(e_m, e_m) = 0$ since g is supported in G.

By Theorem 7 $g = g_C + \sum_{ij} \alpha(i,j) F(i,j)$ where $g_C$ is a closed cycle. Therefore $P(g_c)(e_m) + \sum_{i,j} \alpha(i,j) f(i,j)(e_m) = 0$. $P(g)(e_m) = 0$. But $P(g_c) = 0$ since $g_c$ is a closed cycle. Hence $\sum_{i,j} \alpha(i,j) f(i,j)(e_m) = 0$. But for $i < m$, $\pi_{ij}$ is incident on $G_{m-1}$, that is $e_m$ is not incident on $\pi_{ij}$ and hence $f(i,j)(e_m) = T^n(\pi_{ij})(e_m) = 0$. Therefore $\sum_{i \geq m} \alpha(i,j) f(i,j)(e_m) = 0$ that is $\sum_{i \geq m} f^{(i)}(e_m) = 0$. This proves the theorem.

## 4. The conjectures and their main consequences

**Conjecture 1 :** $\sum_{i>m} f^{(i)}(e_m) = 0$ for $m = 1, 2, \ldots, k-1$.

**Conjecture 2 :** Let $f^{(j+1)}, f^{(j+2)}, \ldots, f^{(k)} = 0$. Then there exists a $g'$ supported in G such that $Pg' = f^{(j)}$.

**Theorem 9 :** If conjecture 1 is true $f^{(m)}(e_m) = 0$ for $m = 1, 2, \ldots, k$.

Proof : Obvious from Theorem 8 and conjecture 1.

**Theorem 10 :** If conjecture 2 is true $f^{(m)}(e_m) = 0$ for $m = 1, 2, \ldots, k$.

Proof : Let $g^{(k)} = g$. We generate $g^{(k-1)}, g^{(k-2)}, \ldots, g^{(1)}$ supported in G with the property that $P g^{(j)} = \sum_{i \leq j} f^{(i)}$ for $j = k-1, k-2, \ldots, 1$. To see this suppose we have generated $g^{(j)}$. Using conjecture 2 there exists $g^{(j)'}$ supported in G such that $P g^{(j)'} = f^{(j)}$. We can take $g^{(j-1)} = g^{(j)} + g^{(j)'}$. Now applying Theorem 8 to $g^{(j)}$ for $m = j$, we get the desired result $f^{(j)}(e_j) = 0$ for $j = k-1, k-2, \ldots, 1$. For $j = k$, we get $f^{(k)}(e_k) = 0$ by using Theorem 8 directly to g for $m = k$. This proves the theorem.

**Theorem 11 :** If either conjecture 1 or conjecture 2 is true, $f^{(m)}$ is a cycle for $m = 1, 2, \ldots, k$.

Proof : We have $f^{(m)} = \sum_{j=1,2,\ldots,d(i)} \alpha(i,j) f(i,j) = \sum_{j=1,2,\ldots,d(i)} \alpha(i,j) T^n(\pi_{mj})$. Because $\{f(i,j)\}_{i,j}$ is a canonical basis $e_m$ is incident on $\pi_{mj}$ for each j. Hence by Theorem 5, $v(f^{(m)}) = f^{(m)}(e_m)$. Thus if conjecture 1 is true, using theorem 9, $v(f^{(m)}) = f^{(m)}(e_m) = 0$, and if conjecture 2 is true we get the same result using Theorem 10. Hence if either conjecture 1 or conjecture 2 is true, $f^{(m)}$ is a cycle.

**Theorem 12 :** Let G be a time-graph of order n which is not hamiltonian. Then any g in $H_P^n$ supported in G is a cycle provided that either conjecture 1 or conjecture 2 is true.

Proof : Suppose either conjecture 1 or conjecture 2 is true. Since there is no $\pi$ in $S_n$ incident on G, d(0) = 0. Thus $P(g) = \sum_{i=1,2,\ldots,k} f^{(i)}$. By Theorem 11, $f^{(i)}$ is a cycle for i=1,2,…,k. Hence P(g) is a cycle. Therefore g is a cycle.

**Theorem 13 :** Let G be a time-graph of order n. then provided conjecture 1 or conjecture 2 is true, G is hamiltonian if and only if there exists a g in $H_P^n$ satisfying

$$v(g) = 1 \qquad (1)$$

and $\qquad$ g(e,e') = 0 for every e in $G^C$ and every e' in E(n). $\qquad$ (2)

Proof : Suppose conjecture 1 or conjecture 2 is true. Any g in $H_P^n$ satisfying (2) is supported in G. If G is not hamiltonian then by Theorem 12, g is a cycle and cannot satisfy (1). Hence if G is not hamiltonian there does not exist any g in $H_P^n$ satisfying both (1) and (2). Conversely if G is hamiltonian then there is a $\pi$ incident on G and then g = $T_P^n(\pi)$ satisfies both (1) and (2). This proves the theorem.

Let $\{g_i\}_{i=1,N}$ with $g_i$ in $T_P^n(S_n)$ be a basis of $H_P^n$. Then for g in $H_P^n$ we can write

$$g = \sum_{i=1,N} \alpha_i g_i$$

Hence by Corollary 4.5, $v(g) = \sum_{i=1,N} \alpha_i$ and for e, e' in E(n), $g(e,e') = \sum_I \alpha_i g_i(e,e')$. Thus from Theorem 13, we get

**Theorem 14 :** Let G be a time-graph of order n and let $\{g_i\}_{i=1,N}$ with $g_i$ in $T_P^n(S_n)$ be a basis of $H_P^n$. Then provided conjecture 1 or conjecture 2 is true G is hamiltonian if and only if the following system of linear equations in the variables $\{\alpha_i\}_{i=1,N}$ has a solution

$$\sum_{i=1,n} \alpha_i = 1$$

$$\sum_{i=1,N} g_i(e,e') \alpha_i = 0 \text{ for every } e \text{ in } G^C \text{ and every } e' \text{ in } E(n). \tag{14.1}$$

## 5. Construction of a basis of $H_P{}^n$ consisting of elements of $T_P{}^n(S_n)$

We assume that $n \geq 3$. For $i$ with $1 \leq i \leq n$ choose any bijection $p_i$ from $\{1,2,\ldots,n-1\}$ to $\{1,2,\ldots,n\} - \{i\}$. Let $S_n{}^i$ be the set of all permutations $\pi$ in $S_n$ with $\pi(1) = i$. For $\pi$ in $S_{n-1}$, define $q_i(\pi) : \{1,2,\ldots n\} \to \{1,2,\ldots n\}$ by

$$q_i(\pi)(1) = i, \text{ and } q_i(\pi)(j+1) = p_i(\pi(j)) \text{ for } 1 \leq j \leq n-1.$$

**Theorem 15 :** For $\pi$ in $S_{n-1}$, $q_i(\pi)$ is in $S_n{}^i$ and $q_i : S_{n-1} \to S_n{}^i$ is a bijection.

Proof : $\pi$ is a bijection from $\{1,2,\ldots,n-1\}$ to $\{1,2,\ldots,n-1\}$ and $p_i$ is a bijection from $\{1,2,\ldots,n-1\}$ to $\{1,2,\ldots,n\} - \{i\}$. Hence $p_i \circ \pi$ is a bijection from $\{1,2,\ldots,n-1\}$ to $\{1,2,\ldots,n\} - \{i\}$. Thus $q_i(\pi)$ restricted to $\{2,\ldots n\}$ is a bijection to $\{1,2,\ldots,n\} - \{i\}$. Since $q_i(\pi)(1) = i$, $q_i(\pi)$ is a bijection from $\{1,2,\ldots,n\}$ to $\{1,2,\ldots,n\}$ and therefore $q_i(\pi)$ is in $S_n{}^i$. Suppose $\pi_1, \pi_2$ be in $S_{n-1}$ such that $\pi_1 \neq \pi_2$. Then for some $j$ with $1 \leq j \leq n-1$, $\pi_1(j) \neq \pi_2(j)$. Therefore since $p_i$ is a bijection $p_i(\pi_1(j)) \neq p_i(\pi_2(j))$. Hence $q_i(\pi_1(j+1)) \neq q_i(\pi_2(j+1))$ and therefore $q_i(\pi_1) \neq q_i(\pi_2)$. Thus $q_i$ is one-to-one and hence is a bijection since $S_{n-1}$ and $S_n{}^i$ have the same number of elements namely $(n-1)!$. This proves the theorem.

Since $q_i$ is a bijection, it has an inverse $q_i{}^{-1}$ from $S_n{}^i$ to $S_{n-1}$.

For $1 \leq i \leq n$, let $E(n,i)$ be the set of edges of $E(n)$ of the form $(i_1, j_1, t_1)$ with $t_1 \geq 2$ and $i_1, j_1 \neq i$. Define the map $r_i$ from $E(n-1)$ to $E(n, i)$ by $r_i((i', j', t')) = (p_i(i'), p_i(j'), t'+1)$.

**Theorem 16 :** $r_i$ is a bijection.

Proof : Since $p_i$ is one-to-one, $r_i$ is also one-to-one. Also since both $E(n-1)$ and $E(n,i)$ have $(n-1)^2 (n-2)$ elements, $r_i$ is a bijection. This proves the theorem.

Since $r_i$ is a bijection, it has an inverse $r_i{}^{-1}$ from $E(n,i)$ to $E(n-1)$.

**Theorem 17 :** For $\pi$ in $S_{n-1}$ and $e$ in $E(n-1)$, $e$ is incident on $\pi$ if and only if $r_i(e)$ is incident on $q_i(\pi)$.

Proof : Let $e = (i', j', t')$. Then $e$ is incident on $\pi$ if and only if $\pi(t') = i'$ and $\pi(t'+1) = j'$, if and only if $p_i(\pi(t')) = p_i(i')$ and $p_i(\pi(t'+1)) = p_i(j')$ since $p_i$ is a bijection. Thus $e$ is incident on $\pi$ if and only if $q_i(\pi)(t'+1) = p_i(i')$ and $q_i(\pi)(t'+2) = p_i(j')$ which is if and only if $(p_i(i'), p_i(j'), t'+1)$ is incident on $q_i(\pi)$. This is if and only if $r_i((i', j', t'))$ is incident on $q_i(\pi)$ that is $r_i(e)$ is incident on $q_i(\pi)$. This completes the proof.

**Corollary 17.1 :** For $\pi$ in $S_n^i$ and $e$ in $E(n,i)$, $e$ is incident on $\pi$ if and only if $r_i^{-1}(e)$ is incident on $q_i^{-1}(\pi)$.

**Corollary 17.2 :** For $\pi$ in $S_{n-1}$ and $e, e'$ in $E(n-1)$

$$T_P^{n-1}(\pi)(e,e') = T_P^n(q_i(\pi))(r_i(e), r_i(e'))$$

**Corollary 17.3 :** For $\pi$ in $S_n^i$ and $e, e'$ in $E(n, i)$

$$T_P^n(\pi)(e, e') = T_P^{n-1}(q_i^{-1}(\pi))(r_i^{-1}(e), r_i^{-1}(e')).$$

**Theorem 18 :** Let $\pi_1, \pi_2, \ldots \pi_N, \pi \in S_{n-1}$ such that

$$T_P^{n-1}(\pi) = \sum_{k=1..N} \alpha_k T_P^{n-1}(\pi_k)$$

Then $\qquad T_P^n(q_i(\pi)) = \sum_{k=1..N} \alpha_k T_P^n(q_i(\pi_k))$

Proof : Since for $g \in H_P^n$, $g(e_1, e_2) = g(e_2, e_1)$, it is sufficient to prove

$$T_P^n(q_i(\pi))(e_1, e_2) = \sum_{k=1..N} \alpha_k T_P^n(q_i(\pi_k))(e_1, e_2) \qquad (1)$$

where $e_1 = (i_1, j_1, t_1)$, $e_2 = (i_2, j_2, t_2)$ and $t_2 \geq t_1$.

Case 1 : $t_1, t_2 \geq 2$. For $t \geq 2$ $q_i(\pi)(t) \neq i$ and hence if $e_1$ or $e_2 \notin E(n, i)$, both sides of (1) are zero. If both $e_1, e_2 \in E(n, i)$ then (1) follows easily from Corollary 17.3.

Case 2 : $t_1 = 1$, $t_2 \geq 2$. Unless $i_1 = i$, $j_1 \neq i$ and $e_2 \in E(n, i)$, both sides of (1) are zero. We now assume $i_1 = i$, $j_1 \neq i$ and $e_2 \in E(n, i)$. Consider $E = \{e \mid e=(j_1, j', 2) \in E(n,i)\}$.

**Lemma 18.1 :** For any $\pi' \in S_{n-1}$

$$T_P^n(q_i(\pi'))(e_1, e_2) = \sum_{e \in E} T_P^n(q_i(\pi'))(e, e_2)$$

Proof : If $e_2$ is not incident on $q_i(\pi')$, both sides are zero. If $e_1$ is not incident on $q_i(\pi')$, the left hand side is zero and $j_1 \neq q_i(\pi')(2)$. Then no $e$ in $E$ is incident on $q_i(\pi')$ and the right hand side is also zero. So finally let us assume that both $e_1, e_2$ are incident on $q_i(\pi')$. Then the left hand side is 1 and $j_1 = q_i(\pi')(2)$. Then exactly one $e$ in $E$ namely for which $j' = q_i(\pi')(3)$ is incident on $q_i(\pi')$ and the right hand side is also 1. This proves the lemma.

Now for the proof of the theorem for case 2,

$$T_P^n(q_i(\pi))(e_1, e_2) = \sum_{e \in E} T_P^n(q_i(\pi))(e, e_2) \quad \text{by Lemma 18.1}$$

$$= \sum_{e \in E} \sum_{k=1..N} T_P^n(q_i(\pi_k))(e, e_2) \quad \text{by Case 1}$$

$$= \sum_{k=1..N} \sum_{e \in E} T_P^n(q_i(\pi_k))(e, e_2)$$

$$= \sum_{k=1..N} T_P^n(q_i(\pi_k))(e_1, e_2) \quad \text{by Lemma 18.1}$$

This proves case 2 of the theorem.

Case 3 : $e_1 = (i_1, j_1, 1)$, $e_2 = (i_2, j_2, 1)$. Unless $i_1 = i_2 = i$ and $j_1 = j_2 = j \neq i$, both $e_1, e_2$ cannot be incident on $q_i(\pi')$ for any $\pi' \in S_{n-1}$ and both sides of (1) are zero. Therefore we take $i_1 = i_2 = i$ and $j_1 = j_2 \neq i$ i.e. $e_1 = e_2$. Consider $E = \{e \mid e = (j_1, j', 2) \in E(n, i)\}$

**Lemma 18.2 :** For any $\pi' \in S_{n-1}$

$$T_P^n(q_i(\pi'))(e_1, e_1) = \sum_{e \in E} T_P^n(q_i(\pi'))(e_1, e).$$

Proof : If $e_1$ is not incident on $q_i(\pi')$ then both sides are zero. So let us assume that $e_1$ is incident on $q_i(\pi')$. Then the left hand side is 1 and $j_1 = q_i(\pi')(2)$. Then only one $e$ inn $E$ namely for which $j' = q_i(\pi')(3)$ is incident on $q_i(\pi')$ and the right hand side is also 1. This proves the lemma.

Now for the proof of the theorem for case 3

$$T_P^n(q_i(\pi)(e_1, e_1) = \sum_{e \in E} T_P^n(q_i(\pi))(e_1, e) \quad \text{by Lemma 18.2}$$

$$= \sum_{e \in E} \sum_{k=1..N} T_P^n(q_i(\pi_k))(e_1, e) \quad \text{by case 2}$$

$$= \sum_{1..N} \sum_{e \in E} T_P^n(q_i(\pi_k))(e_1, e)$$

$$= \sum_{k=1..N} T_P^n(q_i(\pi_k))(e_1, e_1) \quad \text{by Lemma 18.2}$$

This completes the proof of case 3 and of the theorem.

**Theorem 19 :** Let $\{T_P^{n-1}(\pi_k)\}_{k=1..N}$ with $\pi_k$ in $S_{n-1}$ be a basis of $H_P^{n-1}$. Then for any $\pi$ in $S_n^i$, $T_P^n(\pi)$ is in the linear span of $\{T_P^n(q_i(\pi_k))\}_{k=1..N}$.

Proof : Since $\{T_P^{n-1}(\pi_k)\}_{k=1..N}$ is a basis of $H_P^{n-1}$, $T_P^{n-1}(q_i^{-1}(\pi)) = \sum_{k=1..N} \alpha_k T_P^{n-1}(\pi_k)$ for some $\alpha_1, \alpha_2, \ldots \alpha_N$. Hence by Theorem 18 $T_P^n(\pi) = \sum_{k=1..N} \alpha_k T_P^n(q_i(\pi_k))$. This proves the theorem.

**Theorem 20 :** Let $\{T_P^{n-1}(\pi_k)\}_{k=1..N}$ be a basis of $H_P^{n-1}$. Then any g in $H_P^n$ is in the linear span of $\{T_P^n(q_i(\pi_k))\}_{k=1..N;\ 1 \le i \le n}$.

Proof : Take any $\pi$ in $S_n$. Let $\pi(1) = i$. Then $\pi$ is in $S_n^i$. By Theorem 19, $T_P^n(\pi)$ is in the linear span of $\{T_P^n(q_i(\pi_k))\}_{k=1..N}$. Hence $T_P^n(S_n)$ is in the linear span of

$$\{T_P^n(q_i(\pi_k))\}_{k=1..N;\ 1 \le i \le n} .$$

Since $H_P^n$ is the linear span of $T_P^n(S_n)$, this proves the theorem.

Thus we get the following algorithm for the construction of a basis of $H_P^n$ consisting of elements of $T_P^n(S_n)$.

**Algorithm 1 :**

1) If $n \le 3$ construct the required basis directly.

2) Construct a basis $\{T_P^{n-1}(\pi_k)\}_{k=1..N}$ of $H_P^{n-1}$ recursively.

3) B = empty

4) For each i, with $1 \le i \le n$

   a) Take a bijection $p_i$ from $\{1,2, \ldots, n\} - \{i\}$ .

   b) $B = B \cup \{T_P^n(q_i(\pi_k))\}_{k=1..N}$

5) Find a maximal linearly independent subset of B which will be the

   required basis of $H_P^n$.

**Theorem 21 :** Algorithm 1 constructs a basis of $H_P^n$ consisting of elements of $T_P^n(S_n)$ in $O(n^{21})$ time.

Proof : The correctness of the algorithm follows from Theorem 20. For the complexity, step 4(a) takes $O(n)$ time and step 4(b) takes $O(nN)$ time. Since N is $O(n^6)$ the overall complexity of step 4 is $O(n^8)$. Each vector of B is of length $O(N)$, B is of size $O(nN)$. Hence step 5 takes $O(n^2N^3)$ that is $O(n^{20})$ time by standard matrix algorithms. Thus if the complexity of the algorithm is T(n), then

$$T(n) \le K n^{20} + T(n-1) \quad \text{for some } K > 0$$

Hence T(n) is $O(n^{21})$.

## 6. A Polynomial time algorithm for the Hamiltonian Path problem :

Finally we get the following algorithm for the Hamiltonian path problem.

**Algorithm 2 :**

Given a graph G of n vertices to determine whether it has a hamiltonian path.

1) Use the construction of Theorem 1 to obtain $G_T$ which is a time-graph of order n.

2) Use Algorithm 1 to construct a basis $\{g_i = T_{P^n}(\pi_i)\}$ of $H_{P^n}$.

3) Compute $g_i(e, e')$ for every $i = 1,2,\ldots,N$, $e \in G_T^C$ and $e' \in E(n)$.

4) Determine whether the set of linear equations (14.1) has a solution.

5) If there is a solution then declare that G has a hamiltonian path else declare that G does not have a Hamiltonian path.

**Theorem 22 :** If either conjecture 1 or conjecture 2 is true, Algorithm 2 correctly decides whether the given graph has a hamiltonian path in $O(n^{21})$ time.

Proof : The correctness of the algorithm follows from Theorem 1 and Theorem 14. For the complexity, step (1) takes $O(n^3)$ time by Theorem 1. Step (2) takes $O(n^{21})$ time by Theorem 21. Since N is $O(n^6)$, $G_T^C$ and $E(n)$ have $O(n^3)$ elements step (3) takes $O(n^{12})$ time. For step (4), in the set of linear equations (14.1) there are $O(N)$ equations in $O(N)$ variables. Hence this step can be done in $O(N^3)$ that is $O(n^{18})$ time. Thus the overall complexity of the algorithm is $O(n^{21})$. This proves the theorem.

Since the Hamiltonian path problem is NP-complete, this proves that P = NP provided either conjecture 1 or conjecture 2 is true.